\documentclass[iop]{emulateapj}
\usepackage{apjfonts}
\journalinfo{The Astrophysical Journal, 2014, in press}

\shorttitle{Granulation as the Source of Kepler Flicker}
\shortauthors{Cranmer et al.}

\begin{document}

\title{Stellar Granulation as the Source of High-Frequency
Flicker in Kepler Light Curves}

\author{Steven R. Cranmer\altaffilmark{1},
Fabienne A. Bastien\altaffilmark{2},
Keivan G. Stassun\altaffilmark{2,3},
and
Steven H. Saar\altaffilmark{1}}
\altaffiltext{1}{Harvard-Smithsonian Center for Astrophysics,
60 Garden Street, Cambridge, MA 02138,USA}
\altaffiltext{2}{Department of Physics and Astronomy,
Vanderbilt University, 1807 Station B, Nashville, TN 37235, USA}
\altaffiltext{3}{Physics Department, Fisk University,
1000 17th Avenue North, Nashville, TN 37208, USA}

\begin{abstract}
A large fraction of cool, low-mass stars exhibit brightness fluctuations
that arise from a combination of convective granulation,
acoustic oscillations, magnetic activity, and stellar rotation.
Much of the short-timescale variability takes the form of stochastic
noise, whose presence may limit the progress of extrasolar planet
detection and characterization.
In order to lay the groundwork for extracting useful information
from these quasi-random signals, we focus on the origin
of the granulation-driven component of the variability.
We apply existing theoretical scaling relations to predict
the star-integrated variability amplitudes for 508 stars with
photometric light curves measured by the {\em Kepler} mission.
We also derive an empirical correction factor that aims to account
for the suppression of convection in F-dwarf stars with magnetic
activity and shallow convection zones.
So that we can make predictions of specific observational
quantities, we performed Monte Carlo simulations
of granulation light curves using a Lorentzian power spectrum.
These simulations allowed us to reproduce the so-called ``flicker
floor'' (i.e., a lower bound in the relationship between the full
light-curve range and power in short-timescale fluctuations) that
was found in the {\em Kepler} data.
The Monte Carlo model also enabled us to convert the modeled
fluctuation variance into a flicker amplitude directly comparable
with observations.
When the magnetic suppression factor described above is applied,
the model reproduces the observed correlation between stellar
surface gravity and flicker amplitude.
Observationally validated models like these provide new and
complementary evidence for a possible impact of magnetic activity
on the properties of near-surface convection.
\end{abstract}


\section{Introduction}
\label{sec:intro}

The last decade has seen a drastic improvement in the precision of
both stellar observations and the consequent determination of
accurate fundamental parameters for thousands of nearby stars
\citep[e.g.,][]{To10,CM13}.
These improvements have also enabled the discovery and characterization
of extrasolar planets with a range of sizes that now extends down to
less than the radius of the Earth \citep{Bc13}.
However, Sun-like stars are known to exhibit stochastic variations
in both their surface-integrated fluxes (``flicker'') and Doppler
radial velocities (``jitter''),
and these fluctuations may become a source of noise that limits
further progress in planet detection.
We aim to improve our understanding of the physical origins
of the quasi-random variability associated with stellar activity.
This paper focuses on the flicker properties of star-integrated
visible continuum light curves; see \citet{Ss09}, \citet{Bo11},
and \citet{Bv13} for additional studies of radial velocity jitter.

High-resolution observations of the Sun have been important in
allowing us to distinguish clearly between a number of distinct
signals that appear in stellar light curves.
For the Sun, acoustic $p$-mode oscillations dominate at periods
of order 3--5 min.
Granulation (i.e., the photospheric signature of the upper
boundary of the convective zone) dominates at slightly longer
periods extending up to 10--30 min.
Sunspot variability---in combination with the evolution of bright
plage and facular regions--- occurs on timescales of days to months,
with a definite peak at the solar rotation period of $\sim$27 days.
For other stars, the granulation and $p$-mode timescales are
expected to scale with the acoustic cutoff period
\cite[e.g.,][]{Br91} and most spot-related activity (ignoring
impulsive flares associated with magnetic active regions) recurs
with the rotation period.

Although stellar $p$-mode oscillations are increasingly being
used for asteroseismic determinations of stellar masses and radii,
there has been less work done on extracting useful physics from the
granulation signals \citep[see, however,][]{Mi08,Lu09,KM10,Mt11}.
In this paper, we follow up on recent observational results from
the {\em Kepler} mission \citep{Br10} that reported new
correlations between the surface gravities of F, G, and K-type
stars and the short-timescale flicker properties of their light
curves \citep{Bf13}.
We make use of theoretical granulation models developed by
\citet{Sm13a,Sm13b} to produce improved predictions of the
observed light curve properties as a function of the fundamental
stellar parameters.
We also show that the empirical correlation between gravity and
flicker can be interpreted as a direct signature of granulation.

In Section \ref{sec:model} of this paper, we present the details of
the model and we propose an empirical modification to it that
accounts for magnetic suppression of granulation in the hottest
stars of the sample.
Section \ref{sec:monte} describes a Monte Carlo model of the
granulation light curves that we constructed in order to better
understand how the different variability indices of \citet{Bf13}
relate to one another and to the intrinsic granulation power.
Section \ref{sec:flicker} summarizes the results, including
a reproduction of the surface gravity dependence of the short-time
light curve ``flicker.''
Lastly, Section \ref{sec:conc} concludes this paper with a brief
summary, a discussion of some of the wider implications of these
results, and suggestions for future improvements.

\section{The Granulation Model}
\label{sec:model}

\subsection{Empirical Scaling Relations}
\label{sec:model:emp}

\citet{Sm13a,Sm13b} presented theoretical scaling relations for
how granular fluctuations in a star's disk-integrated intensity
should vary as a function of its fundamental parameters.
We apply a slightly modified version of this model below.
The root-mean-square amplitude $\sigma$ of photospheric continuum
intensity variations is specified as a function of
effective temperature ($T_{\rm eff}$), surface gravity ($\log g$),
and stellar mass ($M_{\ast}$).
\citet{Sm13b} derived the following scaling,
\begin{equation}
  \sigma \, = \, 0.039 \, \left[
  \left( \frac{T_{\rm eff}}{T_{\odot}} \right)^{3/4}
  \left( \frac{M_{\odot} \nu_{\odot}}{M_{\ast} \nu_{\rm max}}
  \right)^{1/2} \Phi ({\cal M}_{a})^{2} \right]^{1.03}
  \label{eq:sigma}
\end{equation}
where $\sigma$ is given in units of parts per thousand (ppt)
and $\Phi$ is a dimensionless temperature fluctuation amplitude
that depends on the turbulent Mach number ${\cal M}_{a}$ (see below).
The normalizing constants $T_{\odot} = 5777$ K,
$\log g_{\odot} = 4.438$, and $\nu_{\odot} = 3.106$ mHz are taken
from \citet{Sm13b}.
The peak frequency $\nu_{\rm max}$ of $p$-mode oscillations is
assumed to scale with the acoustic cutoff frequency
\citep[e.g.,][]{Br91,KB95}, with
\begin{equation}
  \nu_{\rm max} \, = \, \nu_{\odot} \frac{g}{g_{\odot}}
  \left( \frac{T_{\odot}}{T_{\rm eff}} \right)^{1/2}  \,\, .
  \label{eq:numax}
\end{equation}
An additional dependence of $\nu_{\rm max}$ on the Mach number
${\cal M}_{a}$ has been proposed \citep[e.g.,][]{Bk11,Bk12},
but we continue to use Equation (\ref{eq:numax}) to retain
continuity with other empirical results from asteroseismology.

Equation (\ref{eq:sigma}) suggests that $\sigma$ is nearly linearly
proportional to the combined quantity
$T_{\rm eff}^{3/4} M_{\ast}^{-1/2} \nu_{\rm max}^{-1/2}$.
This scaling comes from an assumed dependence on the average number
of granules ${\cal N}$ that are distributed across the visible
stellar surface.
\citet{Lu06} and \citet{Sm13a,Sm13b} gave the statistical result that
$\sigma \propto {\cal N}^{-1/2}$, and ${\cal N}$ depends in turn on
the stellar radius $R_{\ast}$ and the characteristic granule size
$\Lambda$.
Some recent sets of convection models \citep[e.g.,][]{Rf04,Mg13}
show that $\Lambda$ is close to being linearly proportional to the
photospheric scale height $H_{\ast} \propto T_{\rm eff}/g$ over
a wide range of stellar parameters.
\citet{Tp13} found a slightly modified correlation
($\Lambda \propto T_{\rm eff}^{1.321} g^{-1.097}$)
when modeling convection for both dwarfs and giants, but we
retain the usual assumption of $\Lambda \propto H_{\ast}$.

\citet{Sm13b} also investigated the dependence of the granular
intensity contrast on the turbulent Mach number ${\cal M}_{a}$
in a series of three-dimensional simulations of photospheric
convection.
We use the \citet{Sm13b} scaling relation for the Mach number,
\begin{equation}
  {\cal M}_{a} \, = \, 0.26
  \left( \frac{T_{\rm eff}}{T_{\odot}} \right)^{2.35}
  \left( \frac{g_{\odot}}{g} \right)^{0.152} \,\, ,
  \label{eq:mach}
\end{equation}
which is then used as input to their parameterized fit for the
root-mean square temperature fluctuation amplitude
\begin{equation}
  \Phi ({\cal M}_{a}) \, = \,
  -0.67 + 8.85 {\cal M}_{a} - 8.73 {\cal M}_{a}^{2} \,\, .
  \label{eq:Phi}
\end{equation}
Note that this fitting formula is strictly applicable only for
the range $0.15 < {\cal M}_{a} < 0.45$.
The above relations describe an initial, unmodified description
of turbulent convection to which we propose a modification in
Section \ref{sec:model:mag}.

\begin{figure}
\epsscale{1.05}
\plotone{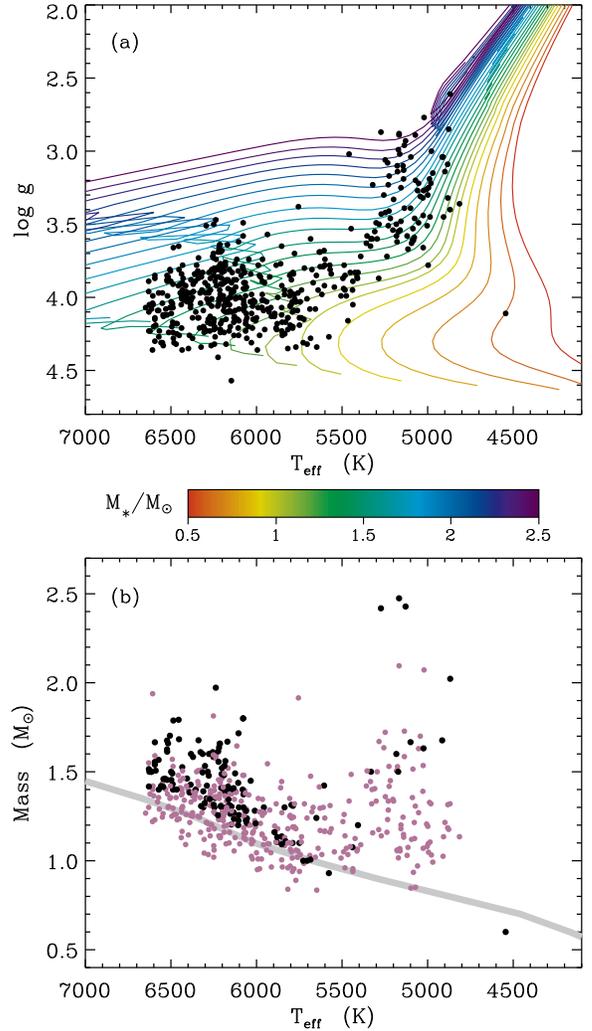}
\caption{(a) Stellar $T_{\rm eff}$ versus $\log g$ for
STARS code evolutionary tracks (solid curves), with masses
labeled by different colors, and for the 508 stars studied by
\citet{Bf13} (solid circles).
(b) Stellar masses from \citet{Ch11b,Ch13} (magenta symbols) and
masses determined from evolutionary tracks (black symbols),
shown alongside an example zero-age main sequence assembled
from models of \citet{Gi00} and \citet{Ek12} (gray curve).
\label{fig01}}
\end{figure}

The 508 {\em Kepler} stars analyzed by \citet{Bf13} all have
measured values of $T_{\rm eff}$ and $\log g$ \citep[see][]{Ch11b,Pn12}.
However, definitive masses for the full set have not yet been
determined.
For a subset of 322 of these stars, we used masses computed
from recent ensemble asteroseismology \citep{Ch13}.
Another 47 of the stars were analyzed in an earlier asteroseismology
study \citep{Ch11b}, and we incorporated those masses as well.
For the remaining 139 stars, we estimated masses by comparing
their measured $T_{\rm eff}$ and $\log g$ values against evolutionary
tracks computed by the Cambridge STARS code \citep{Eg71,Ed08,Ed09}.
These single-star models assumed classic solar abundances
($Z = 0.02$) and did not include mass loss.
Figure \ref{fig01}(a) compares the tracks to the observationally
determined $T_{\rm eff}$ and $\log g$ values of the
{\em Kepler} stars.

Figure \ref{fig01}(b) shows the $T_{\rm eff}$ dependence of the
derived stellar masses.
Both panels in Figure \ref{fig01} make it clear that the {\em Kepler}
sample includes a mix of main-sequence dwarfs, mildly evolved
subgiants, and more highly evolved red giants.
The masses of \citet{Ch11b} and \citet{Ch13} are shown with
magenta symbols.
The black symbols show masses that we determined by
computing a $\chi^{2}$ goodness of fit parameter for each
($T_{\rm eff}$, $\log g$) point along the model tracks and weighting
the final mass determination by $1 / \chi^{2}$ for all models
satisfying $\chi^{2} \leq 10 \min (\chi^{2})$.
We found this to be a less ambiguous process than simple interpolation,
since there are places in Figure \ref{fig01}(a) where the tracks
cross over one another.
This property of the models leads to the existence of multiple
local minima in $\chi^{2}$ space.
Thus, we believe that using a weighted average of the low-$\chi^{2}$
models takes better account of the uncertainties than just selecting
the single model point corresponding to the global minimum in
$\chi^{2}$.

\subsection{Magnetic Inhibition of Convection}
\label{sec:model:mag}

Although F-type stars tend to exhibit lower levels of chromospheric
emission than their cooler G and K counterparts \citep{IF10},
there has been increasing evidence that many of them
have strong enough magnetic activity to suppress the
amplitudes of atmospheric oscillations and granular variability
\citep{Ch11a,Sm13b}.
As $T_{\rm eff}$ increases from 6000 to 7000 K and beyond, the
convection zone shrinks considerably in thickness.
It is suspected that strong-field regions can have a much stronger
inhibitive effect on the correspondingly shallower
granulation of these stars \citep[e.g.,][]{Ca03,Mo08}.
However, the simulations used by \cite{Sm13b} did not contain
this magnetic suppression effect, and thus Equation (\ref{eq:sigma})
was seen to overpredict observed fluctuation amplitudes for the
largest values of $T_{\rm eff}$.

In the absence of a complete theory of magnetic suppression, we
decided on an empirical approach to fold in an additional
$T_{\rm eff}$ dependence to the turbulent Mach number ${\cal M}_a$.
The goal was to leave the predictions of ${\cal M}_a$ unmodified
for the coolest stars and produce a gradual decrease in the
convective velocity field for the hotter, more active stars.
We first used Equation (\ref{eq:mach}) to estimate the Mach
number for each model, then we multiplied it by a dimensionless
suppression factor $S$ given by
\begin{equation}
  S \, = \, \left\{ \begin{array}{ll}
    1 \,\, , & T_{\rm eff} \leq 5400 \, \mbox{K} , \\
    1 / [ 1 + ( T_{\rm eff} - 5400 ) / 1500 ] \,\, , &
      T_{\rm eff} > 5400 \, \mbox{K} . \\
  \end{array} \right.
  \label{eq:Scorr}
\end{equation}
As we describe further in Section \ref{sec:flicker}, this form
for $S$ was selected in order to produce optimal agreement between
the modeled and measured light curve amplitudes for the full range
of F, G, and K stars in the {\em Kepler} sample.
Equation (\ref{eq:Scorr}) is meant to estimate the magnitude of the
suppression effect only for stars with $T_{\rm eff} \lesssim 7000$ K.
There may be additional dependencies on other stellar parameters
for hotter stars with extremely thin surface convection
zones \citep[see also][]{Bc05,Ct09}.

\begin{figure}
\epsscale{1.05}
\plotone{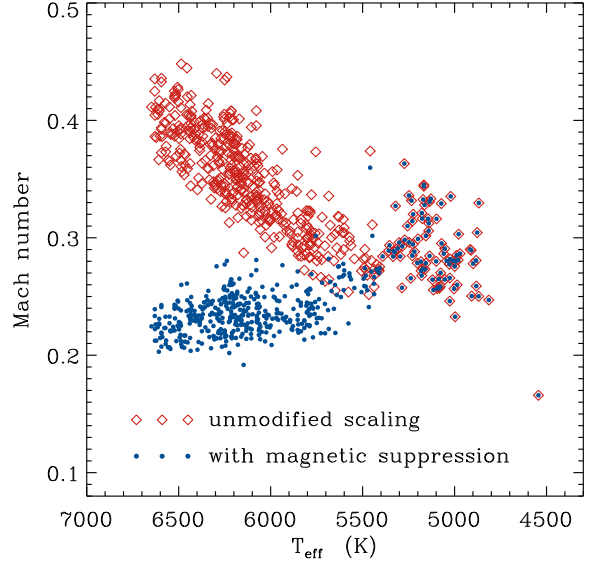}
\caption{Two estimates for the dimensionless convective Mach number
${\cal M}_a$, shown versus $T_{\rm eff}$ for the 508 {\em Kepler} stars.
Red open diamonds show the result of Equation (\ref{eq:mach})
without modification, and blue filled circles show the result of
applying the suppression factor given in Equation (\ref{eq:Scorr}).
\label{fig02}}
\end{figure}

Figure \ref{fig02} shows the impact of applying the correction factor
defined above.
Equations (\ref{eq:mach})--(\ref{eq:Phi}) alone would have predicted
high values of ${\cal M}_{a} \approx 0.4$ and $\Phi \approx 1.5$ for
the hottest stars in the sample.
Our correction gives rise to roughly a factor of 1.8 decrease
in both ${\cal M}_{a}$ and $\Phi$ for the hottest stars.
Note that the revised model prediction for the temperature fluctuation
amplitude produces values that fall closer to the simple approximation
of $\Phi=1$ that was assumed in earlier studies \citep{KB11,Mt11}.

Equations (\ref{eq:sigma})--(\ref{eq:Scorr}) allow us to
compute the root-mean-square amplitude $\sigma$ for each star
in the sample, and for each point along the theoretical
evolutionary tracks.
However, $\sigma$ does not correspond exactly to observational
parameters that are derivable easily from the {\em Kepler}
data.\footnote{%
Although it is possible to compute the full frequency spectrum
from an observed light curve, and process that spectrum to estimate
the values of $\sigma$ and several other granulation properties
\citep[e.g.,][]{Mi08,Mt11}, this procedure is much less
straightforward than the light curve metrics discussed here.}
Following \citet{Ba11}, \citet{Bf13} characterized the
light curves with three independent measures of variability:
\begin{enumerate}
\item
a short-timescale {\em flicker} amplitude ($F_8$) that
corresponds to fluctuations on timescales of 8 hours or less,
\item
the light-curve {\em range} ($R_{\rm var}$), defined as the
difference between the 5\% and 95\% percentile intensities, and
\item
the number of {\em zero crossings} ($Z_{\rm C}$) experienced
by the light curve, smoothed with a 10-hour window, over the full
90 days of the data set.
\end{enumerate}
In this paper we attempt to faithfully reproduce these quantities
as they were defined by \citet{Ba11} and \citet{Bf13}.
However, computing these parameters from the granulation model
requires additional information about the expected frequency
spectrum, which we discuss in the following section.

\section{Statistics of Simulated Light Curves}
\label{sec:monte}

In order to better understand how the various measures of
photometric variability relate to one another,
we constructed Monte Carlo models of light curves
for a range of representative stellar parameters.
We used a Lorentzian form for the granular power spectrum,
\begin{equation}
  {\cal P}(\nu) \, = \,
  \frac{4 \tau_{\rm eff} \sigma^2}
  {1 + (2\pi \tau_{\rm eff} \nu)^{2}}
  \label{eq:Pnu}
\end{equation}
\citep[e.g.,][]{Hv85},
where $\tau_{\rm eff}$ is a characteristic granulation timescale.
The integral of ${\cal P}$ over all frequencies $\nu$ gives the
fluctuation variance $\sigma^{2}$.
However, for these models, we set $\sigma = 1$ in order to
focus on the time-domain structure of the light curves.

The granulation timescales to use in the Monte Carlo models were
obtained by first estimating $\tau_{\rm eff}$ for each of the
observed stars using the scaling relation given by \citet{Sm13b},
\begin{equation}
  \tau_{\rm eff} \, = \, 300 \, \left(
  \frac{\nu_{\odot} {\cal M}_{a \odot}}{\nu_{\rm max} {\cal M}_{a}}
  \right)^{0.98}  \,\,\, \mbox{s} \,\, ,
  \label{eq:taueff}
\end{equation}
where ${\cal M}_{a \odot} = 0.26$.
We used the modified version of ${\cal M}_{a}$ described in
Section \ref{sec:model:mag} as the input quantity to
Equation (\ref{eq:taueff}).
Thus, the estimated values of $\tau_{\rm eff}$ for the {\em Kepler}
stars ranged between 300 and 14,000 s, with a median value of 990 s.
We then created a set of 500 model light curves with a grid of
$\tau_{\rm eff}$ values spread out between 150 and 20,000 s,
slightly wider than the observed range.
Each Monte Carlo light curve was built up from 500 independent
frequency components, where the dimensionless frequency
$2\pi \tau_{\rm eff} \nu$ ranged from 0.1 to 100.1 in constant
steps of 0.2.
Each component was assumed to be a sinusoid with a random phase
and a relative amplitude consistent with ${\cal P}(\nu)$.

\begin{figure*}
\epsscale{1.00}
\plotone{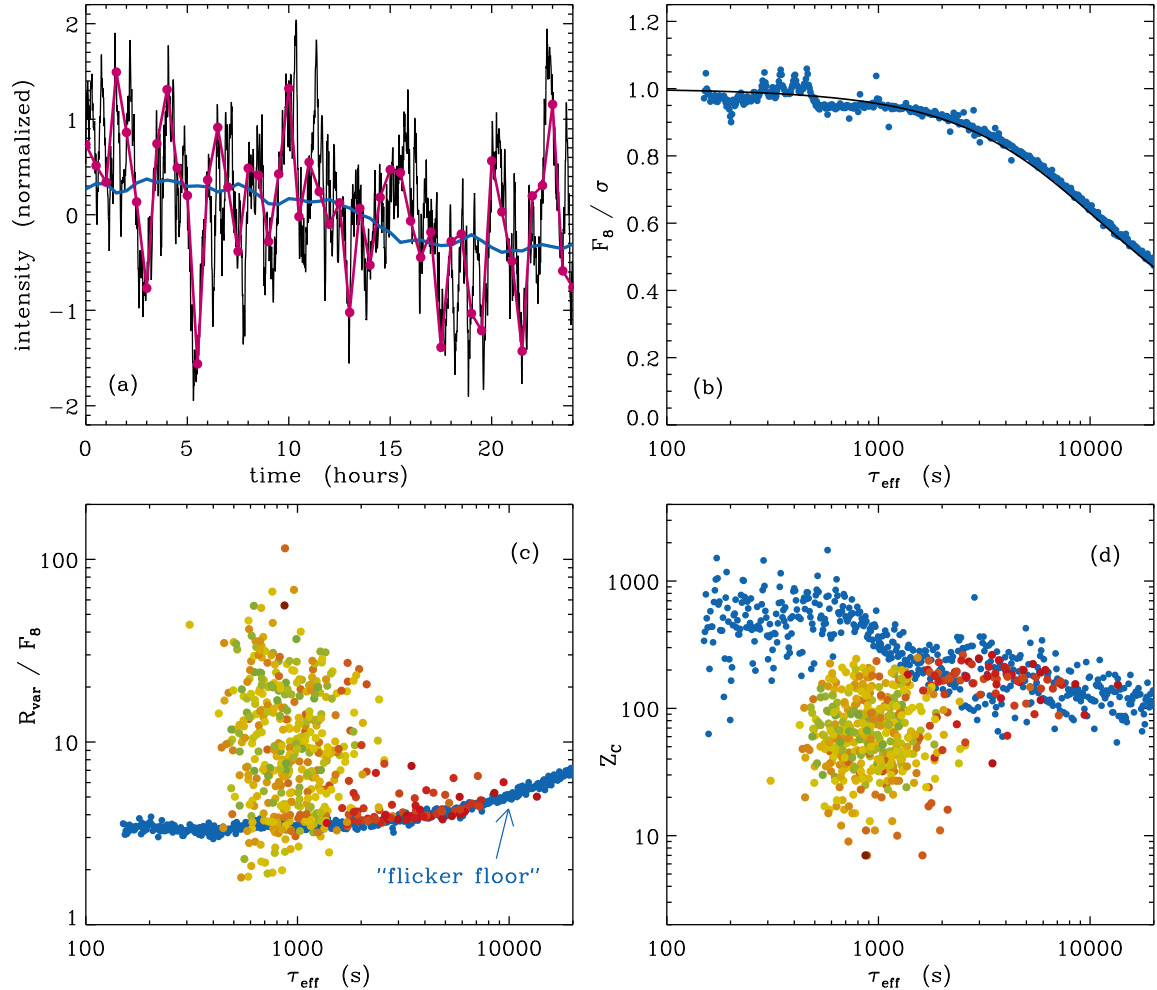}
\caption{(a) Example Monte Carlo light curve, shown at full
resolution (thin black curve) and 30 min {\em Kepler} resolution
(thick magenta curve), along with the 8~hr smoothed version
of the {\em Kepler} light curve that removes nearly all of the
granular variability (blue curve).
(b) $F_{8}/\sigma$ plotted versus $\tau_{\rm eff}$ for the
500 Monte Carlo models (blue points) and as given by
Equation (\ref{eq:F8osig}) (black curve).
(c) $R_{\rm var}/F_8$ versus $\tau_{\rm eff}$ for
the 508 observed stars (with color corresponding to $T_{\rm eff}$
using the scale given in Figure \ref{fig04}) and for the
Monte Carlo models (blue points).
(d) Same as (c), but for $Z_{\rm C}$.
\label{fig03}}
\end{figure*}

For each set of random Monte Carlo variables, we constructed
a full-resolution light curve with a maximum duration of 90 days
(one {\em Kepler} quarter) and a point-to-point time step of 2.5 s.
A corresponding ``observational'' light curve was sampled from it
with a spacing of 30 min---as in the {\em Kepler} low-cadence
data---and it was processed in an identical way as the actual data
to obtain $F_8$, $R_{\rm var}$, and $Z_{\rm C}$.
As an example, Figure \ref{fig03}(a) shows a full resolution model
light curve, its reduced {\em Kepler} sampling, and the 8~hr
smoothing performed to compute $F_8$.
For this model, $\tau_{\rm eff} = 1000$ s, and only a one-day
subset of the light curve is shown.

Figure \ref{fig03}(b) shows the ratio $F_{8}/\sigma$ as a
function of $\tau_{\rm eff}$ for the 500 models.
For the shortest granular timescales (i.e.,
$\tau_{\rm eff} \lesssim 1000$ s), the $F_8$ diagnostic seems to
capture the full granular fluctuation amplitude, and
$F_{8} \approx \sigma$.
Figure \ref{fig03}(b) shows a roughly 10\% random scatter around
the mean ratio $F_{8}/ \sigma = 1$.
We believe this arises from both the random nature of the Monte Carlo
models and the 30-min cadence sampling used to obtain the flicker
amplitude $F_8$.
When the 30-min cadence was replaced by a 10-min cadence, the
spread in $F_{8}/ \sigma$ (which we measured by computing the range
between the 5\% and 95\% percentile values of the ratio, for
$\tau_{\rm eff} < 1000$ s) is reduced from 9.7\% to 6.7\%.
It may be the case that this scatter contributes to the root mean
square deviation of $\sim$0.1 dex that characterizes the spread in
the observed correlation between $F_{8}$ and $\log g$ \citep{Bf13}.

For the largest values of $\tau_{\rm eff}$, a significant
portion of the low-frequency end of the spectrum is excluded
by the 8-hr filtering used to compute $F_8$.
Thus, the ratio $F_{8}/ \sigma$ is reduced to values as low as
$\sim$0.5 at the largest value of $\tau_{\rm eff} =$ 20,000 s.
We can estimate the quantitative impact of the 8-hr filtering
by integrating over the power spectrum (Equation (\ref{eq:Pnu}))
only for frequencies exceeding $\nu_{8} = (8 \, \mbox{hr})^{-1}$.
The result, expressed as a ratio of the relevant root-mean-square
amplitudes, is
\begin{equation}
  \frac{F_8}{\sigma} \, = \, \sqrt{1 - \frac{2}{\pi}
  \tan^{-1} \left( 4 \tau_{\rm eff} \nu_{8} \right)} \,\, .
  \label{eq:F8osig}
\end{equation}
This expression reproduces the Monte Carlo results shown in
Figure \ref{fig03}(b), and we will use it when converting from
$\sigma$ to $F_8$ below.

We also computed the amplitude ratio $R_{\rm var}/F_8$ and
the already dimensionless number $Z_{\rm C}$ from the Monte
Carlo light curves.
\citet{Bf13} noted that the stars that seemed to be least
contaminated by rotating spot activity tend to have the lowest
values of $R_{\rm var}$.
Specifically, when $R_{\rm var}$ is plotted as a function of
$F_{8}$, there is a noticeably sharp ``flicker floor'' below
which no stars seem to appear.
Stars above the floor were interpreted as spot dominated while
those on the floor where interpreted as granulation dominated.
This floor is described approximately
by $R_{\rm var} \approx 3 F_{8}$.
\citet{Bf13} also noted that stars on the flicker floor tend to
have the largest values of $Z_{\rm C}$, a ``zero-crossing ceiling.''

Figure \ref{fig03}(c) compares the modeled $R_{\rm var}/F_8$ 
ratios to the observational data.
Indeed, the modeled ratio sits very near the observed
flicker floor value of 3.
The model ratio also increases slightly for the largest values of
$\tau_{\rm eff}$, and the observed cool, low-gravity giants
in the sample follow this increase as well. 
In the Monte Carlo models, the ratio ratio $R_{\rm var}/ \sigma$
remains roughly constant no matter the value of $\tau_{\rm eff}$,
so the increase in
$R_{\rm var}/F_8$ occurs solely because of the decrease in
$F_{8} / \sigma$ as shown in Figure \ref{fig03}(b).

It is possible to derive an analytic estimate of
$R_{\rm var}/F_{8}$ in the small-$\tau_{\rm eff}$ limit.
For short-timescale granular fluctuations, we showed that
$F_8$ is equivalent to the standard deviation $\sigma$ of the
distribution of intensities that appear in the light curve.
The 30~min cadence observations draw effectively random samples
from this distribution.
If we assume a normal distribution, then the 5\% percentile is
known to be a value that is $1.6449\sigma$ below than mean,
and the 95\% percentile is $1.6449\sigma$ above the mean.
Thus, under these assumptions, $R_{\rm var}/F_{8} = 3.2898.$
For $\tau_{\rm eff} \leq 500$ s, Figure \ref{fig03}(c) shows
that the modeled $R_{\rm var}/F_8$ values are given roughly
by $3.34 \pm 0.16$, which overlaps with the analytic prediction.

Figure \ref{fig03}(d) shows how the Monte Carlo model also seems
to reproduce the ceiling in the observed values of $Z_{\rm C}$
from the {\em Kepler} sample.
As predicted by \citet{Bf13}, stars with light curves dominated
by granulation exhibit the largest number of zero crossings.
Stars with large spots are expected to have light curves dominated
by rotational modulation, which produces substantially lower
frequency variability than does granulation.
For the dwarf stars (i.e., the shortest $\tau_{\rm eff}$
timescales), the Monte Carlo models slightly overestimate the
upper range of observed values of $Z_{\rm C}$.
The cool giants, with $\tau_{\rm eff} \gtrsim 2000$ s, appear
to match the model predictions quite well.
The stars that fall along the $Z_{\rm C}$ ceiling are presumably
(largely) unspotted and therefore granulation dominated.

A simple prediction for the maximum possible number of zero
crossings expected from a noisy, but smoothed, data set would be to
assume that $Z_{\rm C} \approx N/M$, where $N$ is the total number
of data points and $M$ is the width of the boxcar averaging window
used to smooth the data.
For the simulated {\em Kepler} data, $N = 4320$ (corresponding to
90 days with 30-minute cadence) and $M = 20$ (corresponding to a
10 hr window), and thus $N/M = 216$.
However, both the observations and simulations exhibit some data
points with $Z_{\rm C}$ greater than this value.\footnote{%
Observed or simulated values of $Z_{\rm C}$ {\em less} than this
value are easier to understand, since that merely implies a
smoother light curve than one that would vary up and down through
the median value after every measurement.}
To refine this prediction, we performed a set of separate simulations
using a time series of uniformly distributed random numbers
between 0 and 1.
We smoothed them, subtracted their median values,
and counted up the zero crossings in a similar way as was done with
the data.
We varied both $N$ and $M$ in order to determine a robust scaling
relation for $Z_{\rm C}$.
These simulations indicated a mean dependence given roughly by
\begin{equation}
  Z_{\rm C} \, \approx \, \frac{N}{\sqrt{4M}}
\end{equation}
which for the {\em Kepler} data parameters would imply an expectation
of $Z_{\rm C} \approx 480$.
For these parameters, the simulations also showed an approximate
factor of two spread around the mean value.
This is in rough agreement with the simulated light curves shown
in Figure \ref{fig03}(d), which for $\tau_{\rm eff} < 1000$ s
have a mean value of $Z_{\rm C} = 540$.
However, the observed sample of dwarf stars in this part of the
diagram exhibits a maximum $Z_{\rm C}$ that is
about half of that predicted by the Monte Carlo simulations.
This could imply that even the dwarfs with the lowest magnetic
activity still possess some low-level starspot or plage coverage
that reduces $Z_{\rm C}$ via rotational modulation.

\section{Flicker versus Stellar Gravity}
\label{sec:flicker}

\begin{figure}
\epsscale{1.05}
\plotone{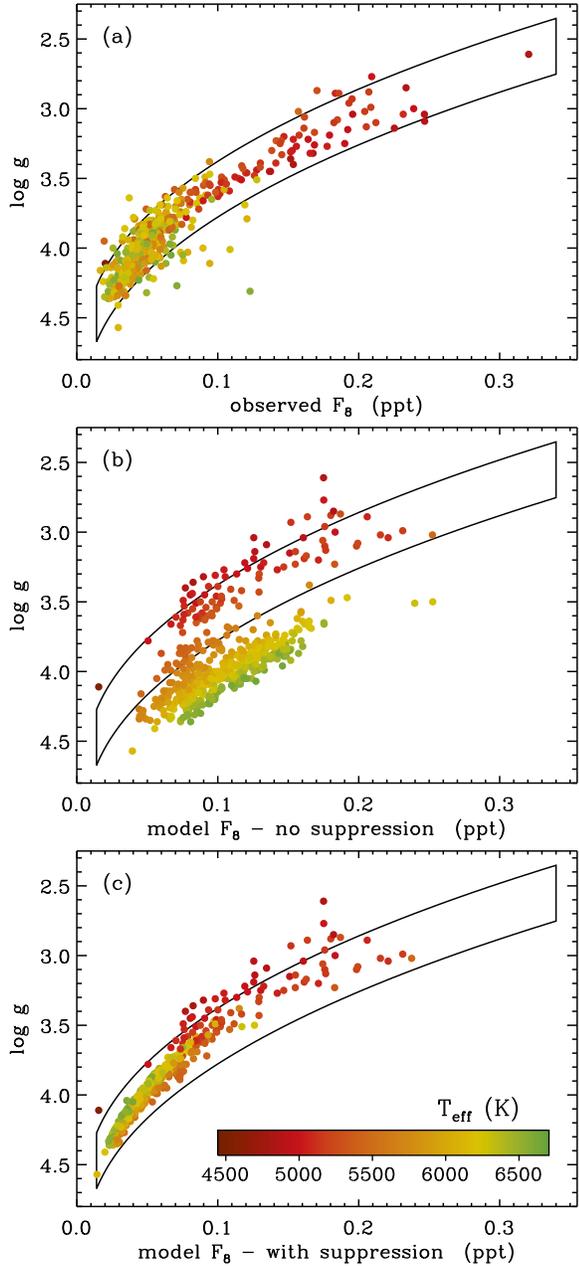}
\caption{Stellar surface gravity dependence of the light curve
flicker amplitude for the {\em Kepler} sample of 508 stars.
(a) Measured values of $F_8$ from \citet{Bf13}.
(b) Modeled $F_8$ computed without magnetic suppression
(i.e., $S=1$).
(c) Modeled $F_8$ computed with Equation (\ref{eq:Scorr}) for $S$.
Symbol color corresponds to $T_{\rm eff}$, and the black solid
outline shows the polynomial fit given by \citet{Bf13} shifted up
and down by factors of $\pm 0.2$ in $\log g$.
\label{fig04}}
\end{figure}

Lastly, we can use the results of the light curve simulations
above to compute model predictions of the flicker amplitude
$F_8$ for the 508 {\em Kepler} stars.
We used Equations (\ref{eq:sigma})--(\ref{eq:Scorr}) to compute
$\sigma$ and Equation (\ref{eq:taueff}) to compute $\tau_{\rm eff}$,
then we applied Equation (\ref{eq:F8osig}) to correct for filtering
to obtain $F_8$.
Figure \ref{fig04} compares the $\log g$ dependence of the
observed flicker parameters to those we computed based on the
above model.
Without including the magnetic suppression effect described in
Section \ref{sec:model:mag}, Figure \ref{fig04}(b) shows that the
flicker amplitudes for stars having $T_{\rm eff} \gtrsim 5700$ K
are distinctly larger than observed.
We applied the suppression effect, as specified
in Equation (\ref{eq:Scorr}), to produce the agreement between the
models and the observations as shown in Figure \ref{fig04}(c).

\begin{figure}
\epsscale{1.05}
\plotone{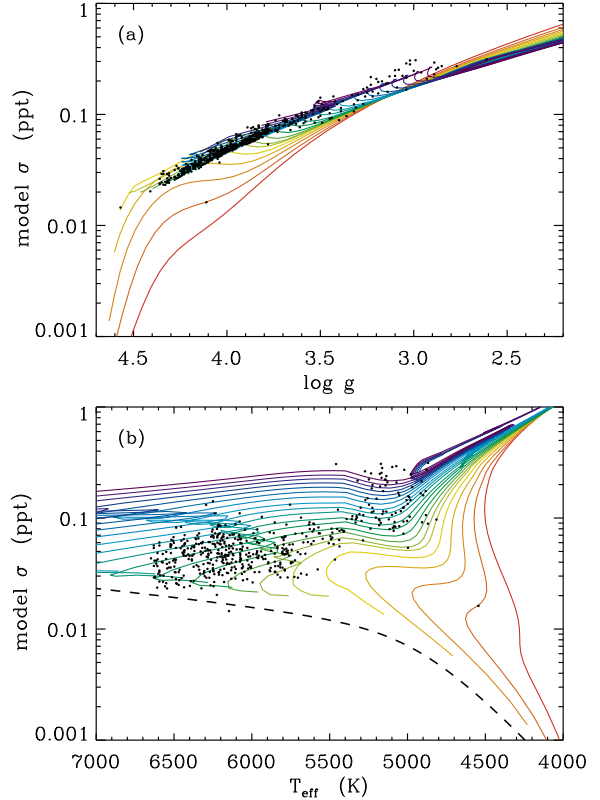}
\caption{Modeled amplitude $\sigma$ shown versus (a) $\log g$
and (b) $T_{\rm eff}$ for the STARS evolutionary tracks.
See Figure \ref{fig01} for the color scale that labels each
track by $M_{\ast}$.
Solid black symbols show the computed $\sigma$ values for
the 508 {\em Kepler} stars, and the dashed curve is
described by Equation (\ref{eq:sigmin}).
\label{fig05}}
\end{figure}

The arrangement of colors in the lower-left part of
Figure \ref{fig04}(c) also suggests there may be a detectable
correlation between the flicker amplitude and $T_{\rm eff}$,
with hotter stars experiencing lower absolute fluctuation levels.
To explore that idea in more detail, Figure \ref{fig05} displays how
$\sigma$ depends separately on $\log g$ and $T_{\rm eff}$ for the
STARS code evolutionary tracks that were shown in Figure \ref{fig01}.
The computed values of $\sigma$ for the {\em Kepler} stars are
also shown in both panels of Figure \ref{fig05} as symbols.
In addition to the tight $\log g$ dependence that is also seen in
Figure \ref{fig04}, there is a slight correlation between $\sigma$
and $T_{\rm eff}$.
However, Figure \ref{fig05}(b) shows that this apparent $T_{\rm eff}$
dependence is likely to be the result of the age spread of the 508
{\em Kepler} stars, and the fact that these stars fall along only
a relatively narrow set of the evolutionary tracks.
As stars with $M_{\ast} \approx 1$--2 $M_{\odot}$ evolve from high
to low $\log g$ and from high to low $T_{\rm eff}$, they
evolve to higher values of $\sigma$.
Thus, the correlation between $\sigma$ and $T_{\rm eff}$ may
disappear for a more heterogeneous set of observed stars with a
broader range of masses.

Along the zero-age main sequence, the dependence of $\sigma$ on
$T_{\rm eff}$ shown in Figure \ref{fig05}(b) can be fit with an
approximate lower-limit amplitude of
\begin{equation}
  \sigma_{\rm min} \, \approx \,
  \frac{0.01 x^{14}}{\sqrt{1 + x^{23}}} \,\, ,
  \label{eq:sigmin}
\end{equation}
where $x = T_{\rm eff}/5000$ K, and we show this relation in
Figure \ref{fig05}(b) with a dashed curve.
We predict very low intensity fluctuations for the coolest main
sequence dwarfs with $\log g > 4.5$ and $T_{\rm eff} < 4500$ K.
For these stars, $\sigma \sim 0.001$ ppt, which would require
extremely precise photometry to measure.
However, these cool dwarfs are also expected to exhibit shorter
$\tau_{\rm eff}$ timescales that could make the flicker signal
more easily measurable.

\section{Discussion and Conclusions}
\label{sec:conc}

The goal of this paper was to apply a model of
cool-star granulation to the photometric variability of a sample
of stars studied by \citet{Bf13}.
An existing model of granular intensity fluctuations was supplemented
by including an ad~hoc, but observationally motivated, correction
factor for the magnetic suppression of convection in the hottest stars.
We also showed how a theoretical frequency spectrum of granular
light curve variability could be used to predict some of the other
properties of the flicker, range, and zero-crossing diagnostics
that were used to analyze the {\em Kepler} data.
The correlation found by \citet{Bf13} between surface gravity and
the flicker amplitude was reproduced with the empirically derived
magnetic suppression factor.
The overall ability of this model to reproduce several different
aspects of the observed light curve variability provides evidence
that short-timescale intensity fluctuations are a good probe of
stellar granulation.

A subsequent goal of this work is to find self-consistent
and accurate ways to predict the properties of stellar light
curve variability, and to use this variability to calibrate
against other methods of determining their fundamental parameters.
Thus, it may be possible to develop the analysis of granular
flicker measurements in a way that augments the results of
asteroseismology and improves the accuracy of, e.g., stellar
mass and radius measurements.
To assist in this process, we provide tabulated data for the
508 {\em Kepler} stars analyzed above, which also includes their
derived masses and predicted values of ${\cal M}_a$, $\sigma$,
and $F_8$.
These data are given as online-only supplemental material
and are hosted, with updates as needed, on the first author's
Web site.\footnote{http://www.cfa.harvard.edu/$\sim$scranmer/}
Packaged with the data is a short code written in the
Interactive Data Language (IDL)\footnote{%
IDL is published by ITT Visual Information Solutions.  There are
also several free implementations with compatible syntax, including
the GNU Data Language (GDL) and the Perl Data Language (PDL).}
that reads the data and reproduces two of the three panels of
Figure \ref{fig04}.

In order to make progress in understanding the properties of
granulation light curves, the models need to be improved in
several key ways.
The scaling relations of \citet{Sm13a,Sm13b} did not include
the effects of varying atmospheric metallicity on $\sigma$. 
\citet{Mg13} presented a grid of three-dimensional granulation
models that were constructed over a range of [Fe/H] values from
--4 to $+$0.5 \citep[see also][]{Lu09,Tb13}.
Metallicity appears to have a significant effect on the
simulation-averaged granulation intensity contrast, but there is
no systematic trend (i.e., in some models, the intensity contrast
goes up with increasing [Fe/H], and in others it goes down).
It would also be useful to combine the granulation model with
accompanying predictions of starspot rotational modulation
\citep[e.g.,][]{Lz12}
and some types of $g$-mode pulsations that can resemble
spot variability \citep{Bn11}.
This could lead the way to predicting the full range of flicker,
range, and zero-crossing parameters for the more active stars
above the flicker floor.

Another source of uncertainty in the models is that our simple
treatment of the magnetic suppression of granulation was assumed
to depend only on $T_{\rm eff}$.
It is clearly desirable to move beyond the empirically guided
(i.e., ad~hoc) correction factor $S(T_{\rm eff})$ that we derived,
but no definite scalings with other stellar properties have emerged.
If strong magnetic fields are really at the root of the velocity
suppression, then for stars with $T_{\rm eff} \lesssim 6500$~K
\citep{BV02} there should be an additional correlation between
velocity suppression and the rotation rate $P_{\rm rot}^{-1}$,
in parallel with well-known relationships between rotation and
high-energy activity \citep[e.g.,][]{Pz03} and rotation and
magnetic flux \citep[e.g.,][]{Rn12}.
A correlation between rotation and the suppression effect, if found,
might explain the added scatter in $F_8$ seen for the higher gravity
stars (Figure \ref{fig04}(a)).
Nevertheless, simulations of the interaction between convection and
inhomogeneous magnetic ``spots'' show that the power in both granular
motions and acoustic $p$-mode oscillations is reduced substantially
when the spot coverage grows larger \citep{Mu73,Ca03,PK07,Mo08}.
Activity-related changes in the convective Mach number may also
affect asteroseismic determinations of quantities like the peak
oscillation frequency \citep{Bk11,Bk12}.
Better models of this interaction may be crucial to improving
predictions of the inflated radii of M dwarfs
\citep[e.g.,][]{MM13,FC13}, and also to understanding how spots are
formed within turbulent convection zones in the first place
\citep{Bd13}.

\acknowledgments

This paper includes data collected by the {\em Kepler} mission.
Funding for {\em Kepler} is provided by the NASA Science Mission
directorate.
The authors gratefully acknowledge William Chaplin and
John Eldridge for making their data available for use in this study.
FAB acknowledges support from a NASA Harriet Jenkins graduate
fellowship.  KGS and FAB acknowledge NSF AST-1009810 and
NSF PAARE AST-0849736.

\end{document}